\documentclass[10pt]{article}
\usepackage[letterpaper]{geometry}
\usepackage{hicss}
\usepackage{times}
\usepackage[none]{hyphenat}
\usepackage{url}
\usepackage{latexsym}
\usepackage{minted}
\usepackage{indentfirst}
\usepackage{graphicx}
\graphicspath{{images/}}
\usepackage[
    style=apa,
  ]{biblatex}
\usepackage{array} 
\usepackage{tabularx}
\usepackage{multirow}
\usepackage{makecell}
\usepackage{graphicx}
\usepackage{enumitem}
\usepackage{amsmath}
\usepackage{amssymb}
\usepackage[locale=FR]{siunitx}
\usepackage{booktabs}
\usepackage{comment}
\usepackage{algorithmic}
\usepackage{cleveref}
\usepackage{subcaption}
\title{An Empirical Framework for Evaluating Semantic Preservation Using Hugging Face}

\author{Nan Jia \\
 CUNY, the Graduate Center \\
 {\underline{njia@gradcenter.cuny.edu}} \\ 
 \And
 Anita Raja \\
 CUNY, Hunter College \\
 CUNY, the Graduate Center \\
 {\underline{ anita.raja@hunter.cuny.edu} } \\ 
\And
 Raffi Khatchadourian \\
 CUNY, Hunter College \\
 CUNY, the Graduate Center \\
 {\underline{ khatchad@hunter.cuny.edu} } \\ 
 }

\date{}

\begin{document}
\maketitle
\begin{abstract}
As machine learning (ML) becomes an integral part of high-autonomy systems, it is critical to ensure the trustworthiness of learning-enabled software systems (LESS). Yet, the nondeterministic and run-time-defined semantics of ML complicate traditional software refactoring. We define semantic preservation in LESS as the property that optimizations of intelligent components do not alter the system’s overall functional behavior. This paper introduces an empirical framework to evaluate semantic preservation in LESS by mining model evolution data from HuggingFace.  We extract commit histories, \textit{Model Cards}, and performance metrics from a large number of models. To establish baselines, we conducted case studies in three domains, tracing performance changes across versions. Our analysis demonstrates how \textit{semantic drift} can be detected via evaluation metrics across commits and reveals common refactoring patterns based on commit message analysis. Although API constraints limited the possibility of estimating a full-scale threshold, our pipeline offers a foundation for defining community-accepted boundaries for semantic preservation. Our contributions include: (1) a large-scale dataset of ML model evolution, curated from 1.7 million Hugging Face entries via a reproducible pipeline using the native HF hub API, (2) a practical pipeline for the evaluation of semantic preservation for a subset of 536 models and 4000+ metrics and (3) empirical case studies illustrating semantic drift in practice. Together, these contributions advance the foundations for more maintainable and trustworthy ML systems.
\end{abstract}

\subsubsection*{Keywords: Refactoring, Semantic Drift, Learning-enabled Software Systems, Hugging Face, Software Evolution}


\section{Introduction}
\label{sec:intro}
In traditional software engineering, behavior-preserving system transformation is a well-understood concept. As first introduced by ~\textcite{opdyke_refactoring_1992}, refactoring in object-oriented programming involves systematically restructuring code without altering its external behavior. However, a learning-enabled software systems (LESS)---where Machine Learning (ML) models and data drive system behavior---refactoring is far more ambiguous. How can we verify that fine-tuning produces a \textit{trustworthy} transformation~\parencite{ao_empirical_2023,jia_reless_2024} that maintains both the system's behavioral integrity~\parencite{tang_empirical_2021} and interpretable decision-making~\parencite{molnar_interpretable_2020}? The uncertainty is especially problematic in high-autonomy domains such as safety-critical infrastructure, finance, and healthcare~\parencite{nahar_product_2024, zhuo_security_2023, hu_if_2022,pan_decomposing_2020}, where ML components must remain reliable and explainable under continuous evolution. 
Yet, despite this need, there is currently no empirical baseline for what counts as a \textit{safe} or \textit{semantics preserving} change when updating models, training data, or documentation. This gap raises risks not only in performance regression but also in trust, reproducibility, and downstream system reliability. 

To mitigate these risks, it is increasingly critical to understand how ML models evolve while retaining their original intent. Unlike traditional software artifacts, ML models are not static---they are rapidly updated through fine-tuning, performance optimization, and documentation updates. Our goal in this paper is to uncover patterns and boundaries of \textit{semantic preservation} during system transformation in widely used pretrained ML models that are hosted on Hugging Face. 

Studying this semantic transformation in deployed LESS is difficult due to proprietary constraints~\parencite{nguyen_open_2025} and dynamic environments~\parencite{david_how_2020, pollano_detecting_2023, hu_if_2022}. 
Fortunately, the Hugging Face platform, often referred to as the “GitHub for ML models~\parencite{pan_empirical_2022, ait_hfcommunity_2023}, offers a uniquely rich and open environment to observe these dynamics at scale. Each model repository hosted on Hugging Face includes not only model weights and configurations but also version-controlled \textit{Model Card}~\parencite{mitchell_model_2019} and commit histories. These artifacts allow researchers to trace intra-repository evolution---how individual models change over time within a single project---not just at a snapshot, but across multiple versions.

In this paper, we present the first empirical study to our knowledge of intra-repository evolution on the Hugging Face (HF) platform by operationally defining semantic preservation via \textbf{metric stability}, which in our work is assessing and visualizing each repository's commit-specific trajectories over temporal changes in their \textit{Model Card}. While our current study defines semantic preservation through the lens of metric stability, we acknowledge that this is a proxy for the broader concept of semantic preservation. The comprehensive definition would include other criteria such as decision boundaries and subgroup performance changes. However, metrics are the most direct evaluation for an engineer to decide whether to integrate a particular ML model into the system without actually executing the ML pipeline, especially in  HF, an open-source platform. Tracking metrics stability and fluctuation suggests intentional tuning or significant structural refactoring. We initially retrieve about a  million open-source real-world  ML repositories, extracting commit logs from sampling, metadata, and evaluation metrics across versions to identify \textit{semantic drift} as measured by how such evolution may affect model behavior and meaning~\parencite{hu_if_2022, pang_robustness_2022} in reported performance. We also analyze the nature of “acceptable" change that preserves semantics using metrics stability analysis. By focusing on documented metric changes in version-controlled model cards, we estimate transformation thresholds that appear consistent with semantic-preserving updates. 

Our study complements, but differs from, prior efforts that characterize platform-wide maintenance behaviors~\parencite{castano_analyzing_2024, ait_hfcommunity_2023}. Although previous studies provide valuable platform-level insights, our analysis is intra-repository, focusing on how individual models evolve and how their evaluation shifts. We argue that semantic preservation must be measurable, comparable, and monitored over time to ensure the reliability of ML systems over their lifetime.

To guide this effort, we pose two research questions:
\begin{itemize}[itemsep=0pt, topsep=0pt, parsep=0pt]
    \item \textbf{RQ1:} How can we empirically measure and define semantic preservation in LESS during code, model, and data evolution?
    \item \textbf{RQ2:} What are the prevalent patterns in ML model update documentation on Hugging Face Hub?
\end{itemize}

To support this analysis, we develop a lightweight extraction pipeline using the HF hub API to retrieve versioned documentation and metadata from model repositories. We identify refactoring-related commits through keyword filtering and analyze changes in evaluation metrics (e.g., \textit{accuracy, f1-score}) over time. We also highlight case studies—image detection, tabular classification, and reinforcement learning—which exhibit a clear pattern of optimized semantic drift across commits.

Together, our findings provide an empirical foundation for understanding semantic preservation in 
ML maintenance. The paper is structured as follows: Section~\ref{sec:related} provides an overview of related research in the field. In Section~\ref{sec:method}, the methodology is detailed, including the creation of a large-scale dataset of ML model evolution, curated from 1.7 million Hugging Face entries via a reproducible pipeline using the native HF hub API, and a practical method for evaluation of semantic-preservation. Experimental results are presented in Section~\ref{sec:experiments}, including how we use our method to explore semantic preservation and empirical case studies demonstrating semantic drift. Section~\ref {sec:conclusion} concludes with remarks and future work.

\section{Related Work}
\label{sec:related}
Previous research on the evolution and upkeep of ML models on Hugging Face~\footnote {\url{https://huggingface.co/models}} has underscored the shifting nature of models and repositories, portraying the platform as an active ecosystem where both technical and social factors drive platform-wide improvements. \textcite{ait_hfcommunity_2023} offered a broad analysis of these aspects by introducing \textit{HFCOMMUNITY}, a tool and dataset that collects and publishes repository and community interactions as a relational database. Likewise, \textcite{castano_analyzing_2024} conducted an in-depth investigation into the progression and maintenance of pre-trained ML models on the HF, highlighting community trends, maintenance patterns, and the increasing demand for structured lifecycle management. Their contribution focuses on mapping high-level trends, repository expansion, and model maintenance patterns, providing a foundation for large-scale empirical studies. While the novelty of these works is in characterizing the broader ecosystem, our work provides a complementary lens by concentrating on intra-repository analysis---specifically, tracking how evaluation metrics vary across model versions to gauge performance consistency and regression risks over time, i.e, whether the semantics (behavior) of LESS has been compromised by such changes.

\begin{figure*}[ht!]
    \centering
    \includegraphics[width=0.9\textwidth]{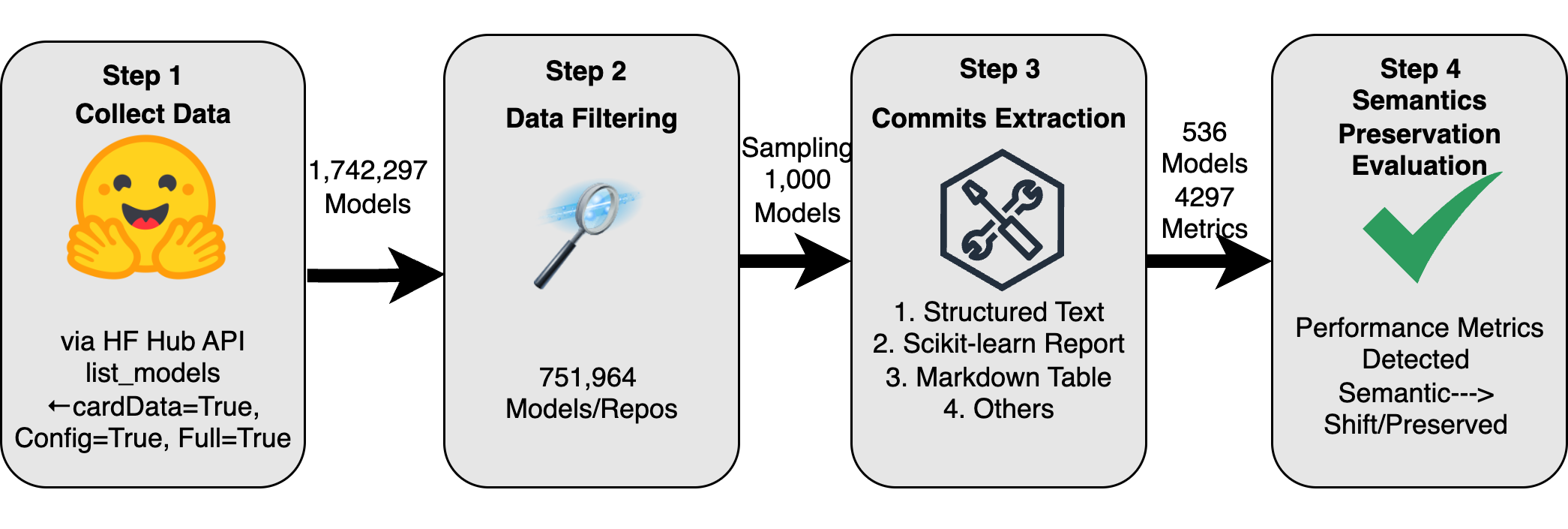}
    \caption{End-to-end pipeline for metadata filtering, metric extraction, and semantic drift detection.}
    \label{fig:pipeline}
    \vspace{-0.5em}
\end{figure*}

Rather than exploring cross-platform evolution, we focus our analysis on the impact of model revision and updates on evaluation stability and trustworthiness, offering a distinct yet synergistic perspective for reliable LESS transformation. Specifically, we look at \textit{ semantics drift}~\parencite{hu_if_2022, pang_robustness_2022}. In the context of systems, developers should practice to preserve the system's intent without influencing end-users. However, semantics in ML systems are nondeterministic due to their statistical nature. If an ML model that previously detected objects for an image task, such as car picture~\parencite{hu_if_2022}, now fails after perturbations such as rotation, this inconsistency suggests a semantic drift, a deviation in the model's behavior that undermines its interpretative consistency. This phenomenon highlights the fundamental gap between human perception and the brittle generalization of some computer vision systems. 

Although techniques such as adversarial training aim to mitigate semantic inconsistencies, structured documentation practices, for example, \textit{Model Cards}~\parencite{mitchell_model_2019}, that accompany trained machine learning models, providing benchmarked evaluations in various conditions, enhance the community's trustworthiness by facilitating the reproducibility and reuse of these intricate models~\parencite{chen_towards_2022}.

Despite growing attention to model evaluation and documentation practices, there remains a lack of empirical understanding of what constitutes safe refactoring across different versions of a model in LESS. While \textit{Model Cards} provide static snapshots of evaluation conditions, they do not systematically track how evaluation metrics evolve over time or respond to code-level or data-level changes. Here, we aim to fill this gap by conducting a comprehensive intra-repository investigation of qualified models on HF to propose empirical thresholds for our goal of semantic preservation.

\section{Methodology/Study Design}
\label{sec:method}
In this section, we address the research questions by detecting semantic drift across versions of ML and DL models by leveraging publicly available metadata and documentation from Hugging Face. To answer RQ1, we curate and filter millions of pre-trained model repositories to retain only non-trivial cases suitable for analysis. For RQ2, we extract functional behavior metrics such as \textit{accuracy} and \textit{f1-score} from versioned \textit{Model Card} via  \textbf{README} documentation, and analyze metric trends across successive commits to evaluate semantic preservation.
We further divide our methodology into multiple stages as captured in Figure~\ref{fig:pipeline}: (1) metadata collection (Step 1) and filtering (Section \ref{subsec:dataset}), (2) performance metric extraction (Section \ref{subsec:perf_metric_extract}) and, (3) evaluation case studies on representative models (Section \ref{subsec:case_study}). The pipeline supports the goal of quantifying practical refactoring thresholds in LESS.

\subsection{Dataset Collection and Metadata Curation}
\label{subsec:dataset}
We begin the pipeline by accessing the HF hub, a Python wrapper that facilitates the retrieval of model and user metadata, through the \texttt{HfApi} interface. As of May 28, 2025, over $1.7$ million models (models are also known as repositories in HF) are hosted on the platform, each containing model ID, configuration files, \textit{Model Cards} (in \texttt{README.md}), and version histories tracked through commit records.  


To ensure the quality and relevance of our analysis, we apply a multi-stage filtering strategy: \\{\textbf{1. Empty model cards:} Repositories with missing or empty \textit{Model Card} files are excluded. Some models claimed to have configuration and \textit{Model Card} information, but the retrieved values are empty objects.\\ \textbf{2. Incomplete configurations:} Models lacking valid architecture fields (e.g., \textit{architectures}) in their \texttt{config.json} files are discarded. The absence of these specifications prevents effective analysis under a transparent software engineering lens, which is essential for studying non-trivial LESS. The data filtering (Step 2) results in \num{751964} models for analysis. We then sampled \num{1000} models from this set with detailed commits history and model card information. Table~\ref{tab:extracted sample} describes one of these data entries.\\ 
\textbf{3. Non-functional keywords filtering in versioned commits:} This involves identifying and managing keywords in documentation that demonstrate indirect contribution to the functional aspects of ML/DL model changes. This filtering is essential for improving the clarity and relevance of the commit messages, ensuring that only significant information is highlighted. Commit extraction  (Step 3), which is further discussed below, results in  536 models and 4297 metrics.}


Compared to prior studies that relied on the snapshot of \num{386006} models in 2023~\parencite{castano_analyzing_2024} and \num{681682}~\parencite{ait_hfcommunity_2023} models in 2024~\footnote{\url{https://som-research.github.io/HFCommunity/download.html}}, our experimental investigation uses a significantly larger and more current dataset of over {\bf $1.7$ million entries}. Attempts to replicate prior filtering strategies yielded limited insights into non-functional behavior. This is largely because those studies focused on global characteristics of model maintenance and evolution, such as popularity and community-level metadata (e.g., likes and downloads) on Hugging Face. These efforts provide valuable ecosystem-wide insights, but our interest lies in the fine-grained evolution within individual repositories, particularly their commit histories and documentation updates related to semantic preservation. To illustrate this, in contrast to previous studies~\parencite{ait_hfcommunity_2023, castano_analyzing_2024}, we design a streamlined pipeline centered on the native HF API endpoint \texttt{list\_repo\_commits}, which eliminates the need for external crawling tools for commits. 

Following established analytical procedures~\parencite{castano_analyzing_2024, noei_detecting_2025}, we further analyze the subset of 751k models to investigate semantic preservation trends across their version histories.

\subsection{Performance Metric Extraction}
\label{subsec:perf_metric_extract}
To analyze how non-functional properties—especially \textit{perfective} commits~\parencite{sarwar_multi-label_2020}—evolve across model versions, we extract key evaluation metrics such as \texttt{refactor}, \texttt{optimize}, \texttt{update}, as well as \texttt{chore}, \texttt{style}, \texttt{test}, and \texttt{security}, each of which may signal authors' intent to modify model implementation while preserving functionality. These activities could imply code restructuring, performance or maintainability improvements, and removal of unused elements. While \texttt{test} is traditionally preventive, we include it following \textcite{sarwar_multi-label_2020}'s rationale, since its presence can indicate broader maintainability improvements. We distinguish between functional and non-functional code changes based on commit message keywords. A functional change is the introduction of a new feature, typically indicated by keywords such as “\texttt{feat}". In contrast, a nonfunctional change is a refactoring or improvement to the codebase that does not alter its user-facing capabilities. For example, a commit message titled “refactor input preprocessing pipeline" uses the keyword “\texttt{refactor}" to denote an intentional, non-breaking change aimed at improving the code's internal structure and maintainability. More details can be found in our public repository.~\footnote{\url{https://github.com/NanJ90/reless_semantic_prev}}


During this phase, we sampled a subset of our filtered dataset to examine how refactoring-related keywords are distributed in practice. To focus the analysis on non-functional improvement, we adopt a simplified version of the maintenance taxonomy proposed by~\textcite{sarwar_multi-label_2020, castano_analyzing_2024}, concentrating on the \textit{perfective} (non-functional improvements) category. The other two categories—\textit{corrective} (bug fixes) and \textit{adaptive} (new feature additions)—typically alter the system’s functional behavior and are thus outside the scope of our semantic preservation analysis.

We initially used a rule-based extraction pipeline for its efficiency and reproducibility. This process involved parsing structured JSON/YAML, scikit-learn style reports, and Markdown tables using a combination of regex patterns and a markdown parser. For unstructured documentation, a fallback regex was used to capture metrics. All extracted numeric values were then normalized for comparability. However, this approach proved to be inefficient on the diverse and unstructured documentation formats found in the dataset. As a result, we transitioned to using a GPT-5-involved API call for more robust and accurate extraction. This allows us to process the full text of documentation and generate a structured summary of performance metrics, overcoming the limitations of static regex patterns.

After collecting this data, we conducted a performance metrics analysis. We define \textit{semantic preservation} of consistent predictive performance across versions of models, as measured by \textit{accuracy, precision, f1-score, and loss} defined in equation~\ref{eq:accuracy}, equation~\ref{eq:precision}, equation~\ref{eq:f1_score}, and equation~\ref{eq:cross_entropy_loss} respectively. This operationalization follows prior work~\parencite{pan_decomposing_2020, pan_decomposing_2021} which equates semantic meaning in ML/DL models to the mapping between inputs and labels. When these performance metrics remain unchanged across model commits, we consider the underlying semantic function preserved. 
Figure~\ref{fig:faid_drift_plot} later visualizes how such metrics evolve over time for case study models.
\vspace{-1em}
\begin{equation}
    \text{Accuracy} = \frac{\text{True Positives} + \text{True Negatives}}{\text{Total Population}}
    \label{eq:accuracy}
\end{equation}
\vspace{-0.5em}
\begin{equation}
   \text{Precision} = \frac{\text{True Positives}}{\text{True Positives} + \text{False Positives}}
   \label{eq:precision}
\end{equation}
\vspace{-0.5em}
\begin{equation}
   \text{F1-Score} = 2 \times \frac{\text{Precision} \times \text{Recall}}{\text{Precision} + \text{Recall}}
   \label{eq:f1_score}
\end{equation}
\vspace{-0.5em}
\begin{equation}
    L(y, \hat{y}) = - \sum_{i=1}^{N} y_i \log(\hat{y}_i)
    \label{eq:cross_entropy_loss}
\end{equation}
\vspace{-1.2em}

\section{Experiments}
\label{sec:experiments}

In this section, we present an empirical analysis of semantic preservation in evolving ML repositories hosted on HF. This section is organized into: 1) exploratory insights into the scale and types of repositories in our dataset, 
2) commit-level analysis of refactoring intent and metric stability analysis and statistical analysis (Section~\ref{subse:perf_trend_and statistic}), and 3) detailed case studies in three domains—image detection, tabular classification, and reinforcement learning—exemplars demonstrating the generalizability of our method.

\subsection{Overview of Datasets Mining}


To contextualize our findings, we conducted exploratory data analysis (EDA) on a filtered corpus of \num{1742297} repositories. 
Transformer-based models for text generation and classification dominate the landscape. As observed in our dataset, a small number of models account for the vast majority of downloads and activity—a characteristic long-tail distribution. This long-tail distribution reinforces the importance of exploring uncovered refactoring patterns and performance trends in intra-repository evolution.


\subsection{Refactoring Patterns in Commit Messages}

To further demonstrate our semantic drift analysis, we examine the distribution of refactoring-related keywords in commit messages across sampled repositories. 

Table~\ref{tab:extracted sample} presents one sample data entry out of the $1,000$ sampled commits. \texttt{id} is the model or repository ID to locate corresponding metadata and temporal information by calling \texttt{list\_models}. Then, we called the \texttt{list\_repo\_commits} to retrieve the commit history for each model based on its ID. This sampling approach, while limited by API constraints, provided significant insights into how ML models document their changes.
\begin{table*}[!htbp]
\small
\centering
\begin{tabularx}{\textwidth}{@{} 
    >{\raggedright\arraybackslash}p{2cm}  
    >{\raggedright\arraybackslash}p{2cm}  
    >{\raggedright\arraybackslash}p{2cm}  
    >{\raggedright\arraybackslash}p{2cm}  
    >{\raggedright\arraybackslash}p{1.2cm}  
    >{\centering\arraybackslash}p{0.1cm}      
    >{\raggedright\arraybackslash}X         
    >{\raggedright\arraybackslash}X         
@{}}
\toprule
\textbf{id} & \textbf{author} & \textbf{sha} & \textbf{created\_at} & \textbf{last\linebreak[1]\_modified} & \textbf{\dots} & \textbf{config} & \textbf{commit} \\
\midrule
Falconsai\slash nsfw\linebreak[1]image\_detection
 & Falconsai & 0436\ldots069 & 2023-10-13 & 2025-04-06 & \texttt{\dots} & 
\makecell[l]{\texttt{\{"arch":}\\ \texttt{["ViT\linebreak[1]For"]\}}}
 & \makecell[l]{\texttt{GitCommitInfo(} \\ \texttt{commit\_id=\dots)}} \\
\bottomrule
\end{tabularx}
\caption{Attributes in the repository data.}
\label{tab:extracted sample}
\end{table*}

\begin{figure}[htbp]
    \centering   \includegraphics[width=0.9\linewidth]{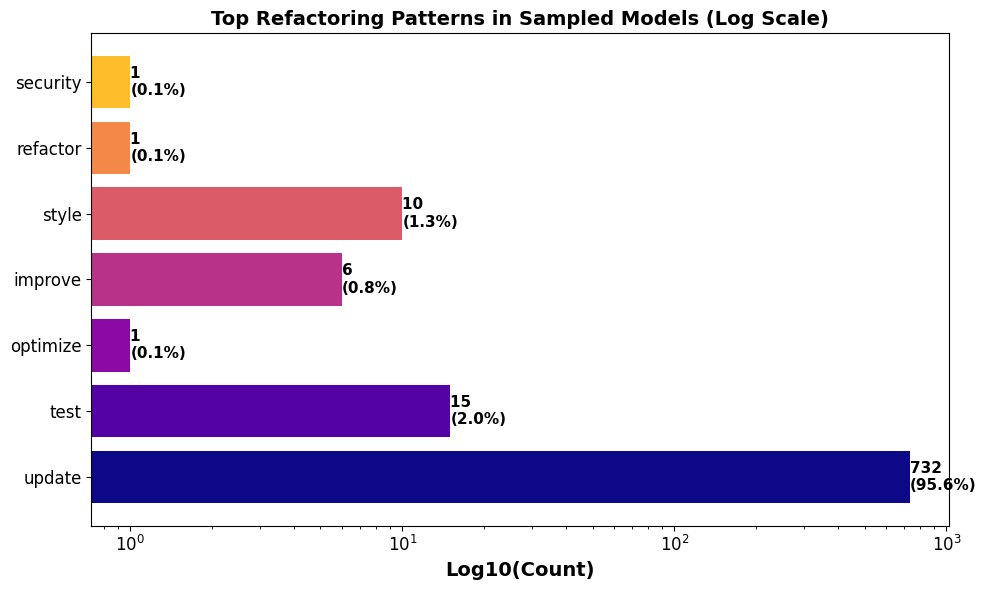}
    \caption{Refactoring Patterns of sampling data---log-scale chart showing the distribution of top refactoring-related keywords in sampled HF model repositories. The vast majority of relevant commit messages use the term “update", while explicit mentions of “refactor", “optimized", or “security" are rare, highlighting an imbalance in how NF changes are documented.}
    \label{fig:refactoring patterns with sampling}
\end{figure}

Figure~\ref{fig:refactoring patterns with sampling} presents a log-scale frequency chart of commonly used terms from sampled data. The vast majority of commits ($95.6$\%) include the generic keyword \texttt{update}. Next, \texttt{test}, \texttt{style} and \texttt{improve} represented as $2$\%, $1.3$\% and $0.8$\%. However, explicit mentions of non-functional maintenance activities such as \texttt{refactor}, \texttt{optimize}, or \texttt{security} are relatively rare at $0.1$\%, $0.1$\%, and $0.1$\%. We found \textit{title} attribute from \texttt{list\_repo\_commits} carries useful guidance about this refactoring and not every non-trivial model's commit comes with commit messages. 
This suggests that refactoring behaviors are often under-documented or embedded within general-purpose commit labels. 

Another observation is that functional changes may consistently utilize the same keyword \texttt{update}. For instance, developers documented the commit title as \texttt{Updated bug in TensorFlow usage code}~\parencite{castano_analyzing_2024}. Future research should focus on the development of more sophisticated information distillation tools ~\parencite{sarwar_multi-label_2020, huang_fewer_2024}. 
Due to inconsistent commit granularity and natural language variation, we then delve into metric-based evaluation to confirm whether such changes indeed preserve or improve semantic functionality.

\subsection{Performance Trends Across Versions}
\label{subse:perf_trend_and statistic}
To investigate semantic preservation on repositories, we applied our metric extraction pipeline (Section~\ref{subsec:perf_metric_extract}) across version histories. 

\begin{figure}[!hb]
    \centering
    \begin{subfigure}[b]{0.9\linewidth}
    \centering
    \includegraphics[width=\linewidth]{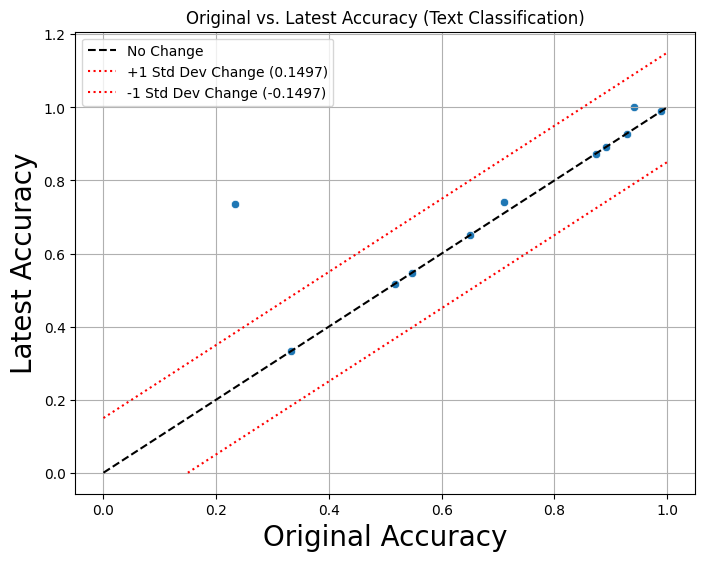}
    \label{acc_text_class_plot}
    \caption{}
    \end{subfigure}
    \hfill
    \begin{subfigure}[b]{0.9\linewidth}
        \centering
        \includegraphics[width=\linewidth]{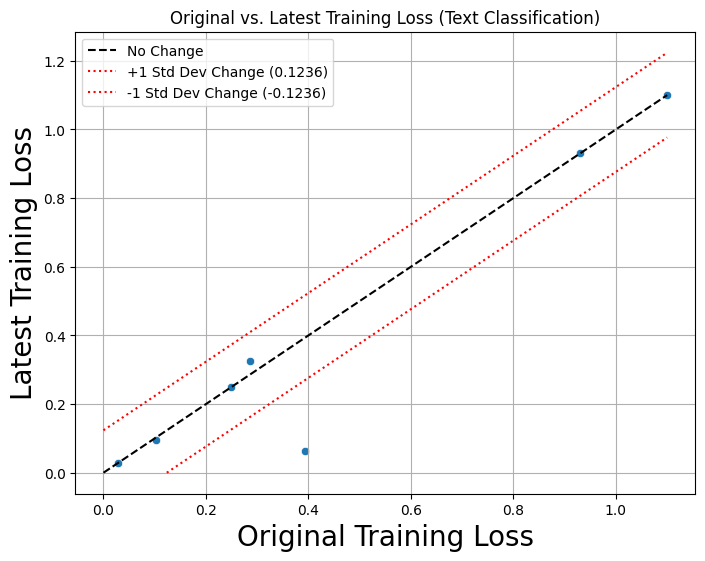}
        \label{training_loss_text_class_plot}
        \caption{}
    \end{subfigure}
    \caption{“Acceptable" change in Accuracy and Training Loss. (a) and (b) illustrate the boundaries of “acceptable" change in accuracy and training loss in a classification task. These plots show that the majority of models exhibit semantic preservation (within $\pm0.15$ for accuracy or $\pm0.13$ for training loss), which aligns with our statistical findings. The CIs, which include zero, suggest that the semantics are preserved while observed changes are not statistically significant.}
    \label{fig:placeholder}
\end{figure}

By plotting these metrics over time for individual models, we observe three recurring patterns. \textbf{Type 1: Optimized Drift} Metrics improve steadily across commits. Out of 195 reported metric changes, 74 are classified within this category; however, only 11 models out of 123 meet the criteria for inclusion in this category. \textbf{Type 2: Semantic Preservation} No significant change despite updates. We observe that the second most of the models fall into this category. For instance, \texttt{Falconsai:nsfw\_image\_detection} repository and \texttt{bert-base-uncased} reported the same classification performance (\textit{accuracy:} 98\%, and \textit{loss} 7.5\%)  via multiple commits ranging from November 2023 to May 2025. \texttt{google-bert/bert-base-uncased} (\textit{accuracy:} 79.6\%)~\parencite{devlin_bert_2019}, from January 2021 to January 2024, remain unchanged, suggesting semantic preservation. Task-based analysis also aligned with this observation. Figure~\ref{fig:placeholder} shows that, in the text classification task, 8 out of 11 models maintain the same accuracy performance. \textbf{Type 3: Performance Degradation} Although authors keep testing and experiments refactoring for better system design and model performance, the performance is not always improved. For instance, Figure~\ref{fig:tabular-drift} presents a fluctuating trajectory of accuracy for a classification task with tabular data. \textbf{Type 4}: Unverifiable Cases: Lack of consistent metrics precludes judgment. 34 out of 123 models fall into this category.

These patterns are analyzed in  Section~\ref{subsec:case_study} as part of the case studies.

There are 89 models that report at least one semantic drift. The percentage of models with one report drift is 16.6\%. A key limitation of our study, as detailed in Table~\ref{tab:ml_metrics} is that the confidence intervals (CIs) for our primary outcomes included zero, indicating the semantic drift is equal to zero, which means the statistics are insignificant. This result may stem from several factors. The study may have been underpowered due to a small sample size or high variability in the data, which reduces the precision of our estimates. This is mainly because each model has a specified dataset and task, evaluated by different metrics.
\begin{table*}[htbp]
    \small
    \centering
    \begin{tabular}{p{3cm}c p{2cm} l r r r r}
        \hline
        Task & \# of Model & Metric & Mean Change & Std Dev Change& 95\% CI Lower & 95\% CI Upper \\
        \hline
        \multirow{4}{*}{text-classification} & \multirow{4}{*}{24} & Accuracy & 0.053366 & 0.149748 & -0.047236 & 0.153968 \\
        & & F1-score & 0.011736 & 0.020821 & -0.004269 & 0.027741 \\
        & & Training loss & -0.052229 & 0.123563 & -0.166505 & 0.062048 \\
        & & Validation (evaluation) loss & 
        0.026800 & 0.188588 & -0.087163 & 0.1407 \\
        \multirow{3}{*}{text-generation} & \multirow{3}{*}{16} & Accuracy & 0.077700 & 0.141418 & -0.097893 & 0.253293 \\
        & & Mean reward & 159.453333 & 217.935605 & -381.928723 & NaN \\
        & & Validation (evaluation) loss & 0.203467 & 0.325495 & -0.138119 & 0.545053 \\
        image-classification & 9 & Accuracy & 0.080131 & 0.133039 & -0.085059 & 0.245320 \\ 
        tabular-classification & 2 & Accuracy & -0.006974 & NaN & NaN & NaN \\
        \hline
    \end{tabular}
    \caption{Performance metrics for four out of the 89 Models we analyzed are shown. The 95\% confidence intervals (CIs) for the average change in performance metrics across different model task categories is also shown. The statistics reveal: 1. \textbf{Intervals containing zero} suggest that the observed average change in the metric is not statistically significant. This is mainly because each model has a specified dataset and task, evaluated by
different metrics.  2. \textbf{Wide Intervals} (e.g., [-381.93, 700.84] for mean reward under text-generation) reflect high variability in the reported changes among models within a task category and/or a small sample size, reducing the precision of the estimate. 3. \textbf{\texttt{NaN} entries} indicate that there were fewer than two models reporting a change for a given metric, making the calculation of a CI statistically infeasible and highlighting data sparsity in certain categories.}
    \label{tab:ml_metrics}
\end{table*}
\subsection{Case Studies in Image, Tabular Data classification and Reinforcement Learning}
\label{subsec:case_study}
We selected Fairface Age Image Detection (FAID)~\parencite{dmytro_iakubovskyi_fairface_2024} as a representative case due to its frequent commits and meaningful version evolution, along with tabular data classification, NYC SQF ARR MLP~\parencite{pereiraPppereira3NYC_SQF_ARR_MLPHugging2024}, and reinforcement learning, Atari Icehockey Superhuman~\parencite{stammersMattStammersAppoatari_icehockeysuperhumanMain2023}. 

FAID is a Vision Transformer (ViT) variant fine-tuned for fair facial recognition. It underwent 13 commits over a 10-day period, from December 6 to December 15, 2024. Throughout these iterations, accuracy improved from 18\% to 53.85\%, and finally to 58.09\%, yielding a total performance delta of 40.09\%. This suggests a gradual refinement process rather than architecture-level changes. The original baseline—\texttt{google\slash vit\linebreak[1]base-patch16-224-in21k}—underperformed due to domain mismatch, but fine-tuning parameters on the FairFace dataset enabled substantial gains. 

\begin{figure*}[!ht]
\centering
        \begin{subfigure}[b]{0.8\textwidth}
            \centering
            \includegraphics[width=\textwidth, height=0.3\textwidth]{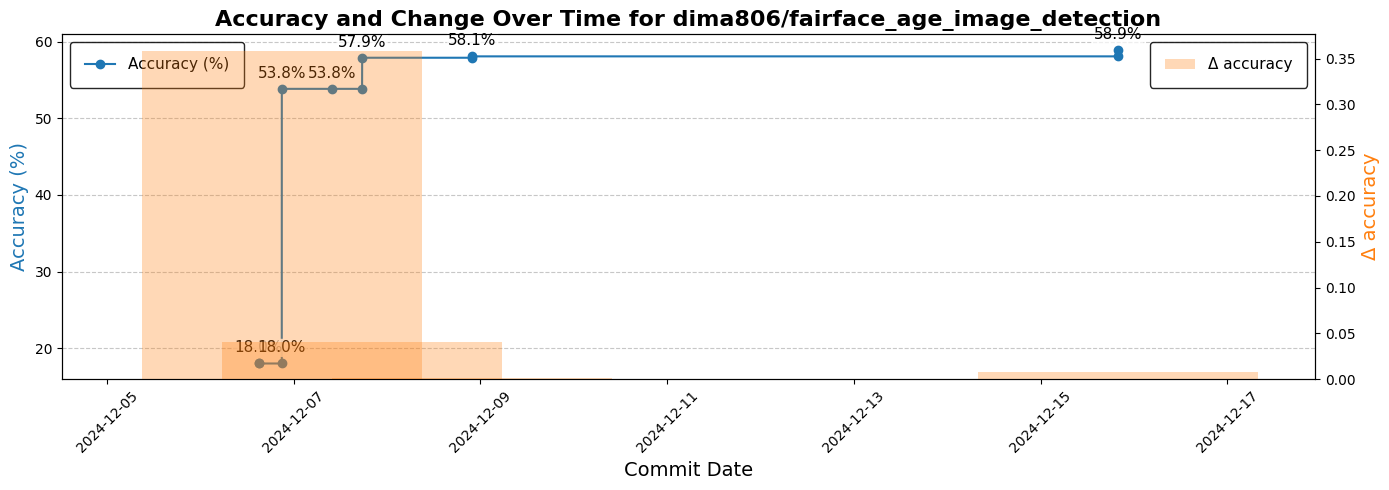}
            \caption{}
            \label{fig:accuracy_fairface}
        \end{subfigure}
        \hfill
                
        \begin{subfigure}[b]{0.8\textwidth}
            \centering
            \includegraphics[width=\textwidth, height=0.3\textwidth]{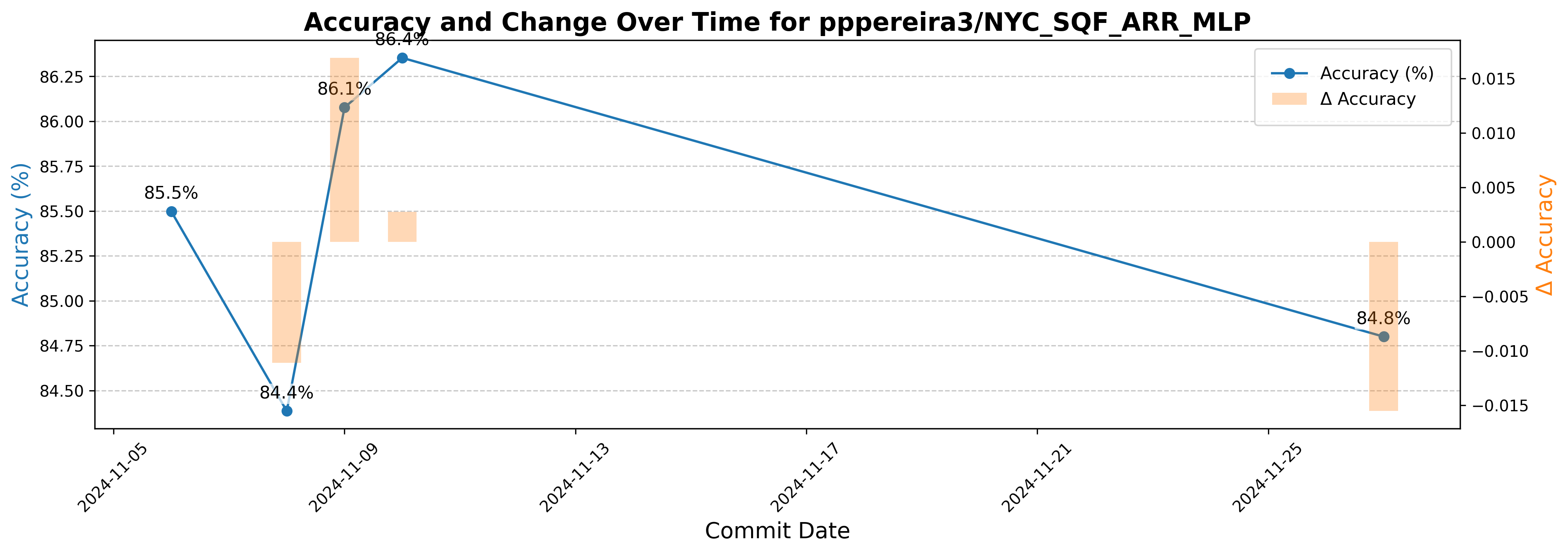}
            \caption{}
            \label{fig:tabular-drift}
        \end{subfigure}
    \hfill
        \begin{subfigure}[b]{0.8\textwidth}
            \centering
            \includegraphics[width=\textwidth, height=0.3\textwidth]{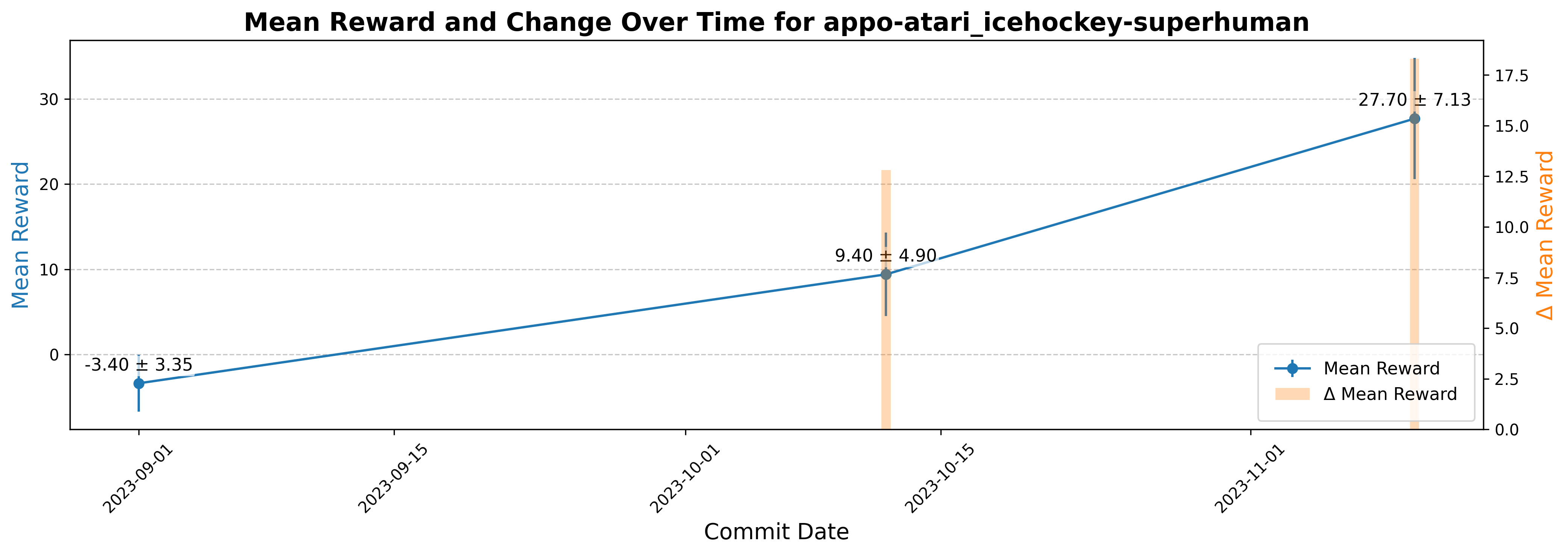}
            \caption{}
            \label{fig:rl-drift}
        \end{subfigure}
        \caption{Semantic Drifts in Different Tasks. (a) shows the accuracy of an image classification model. The initial accuracy of the pre-trained model on the target domain was 18\%. After fine-tuning with new data, the accuracy improved to 58.9\%. (b) displays the accuracy of a tabular data classification. The model's performance fluctuated, starting at 85.5\% on 11/5/2024 and decreasing to 84.8\% on 11/25/2024. (c) presents the mean reward for a reinforcement learning task. The model's performance improved significantly, with a total gain of $31.1\pm 7.88$. The shaded areas ($\Delta$ accuracy) indicate the uncertainty for each metric's value, reflecting intentional, non-breaking refactorings for task-specific optimization. Accuracy is used as the primary metric for classification tasks due to space limitations. Precision, recall, and f1-score followed similar trends }
        \label{fig:faid_drift_plot}
        \vspace{-0.5em}
\end{figure*}

Furthermore, a tabular data classification repository, \texttt{pppereira3/NYC\_SQF\_ARR\_MLP}, is a model training on tacular data related to stop-and-frisk incidents in NYC. After a set of refactoring on feature sets, model parameters, and training data, it became more aggressive in predicting the positive class, which increased recall (70.7\% to 73.4\%) by capturing more true positives but simultaneously decreased precision (81.9\% to 78\%), f1-score (77.7\% to 75.6\%), and overall accuracy (86.4\% to 84.8\%) as shown in Figure~\ref{fig:tabular-drift} due to a surge in false positives. Another is a reinforcement learning (RL) task that focuses on enhancing the performance of high sample throughput APPO RL models in Atari environment, and the measurement is the mean reward. The initial performance is  $-3.4$ ($\pm3.35$), and after a few refactoring, the performance increased to $27.7$ ($\pm7.13$). This is depicted in Figure~\ref{fig:rl-drift}. In essence, the increase in mean reward is attributed to the maintenance and optimization of the model for its specific task in the Atari environment rather than a major architecture change. These additional cases help to illustrate that our pipeline can be applied to a variety of tasks and that semantic preservation should also be considered from fairness and robustness perspectives.

These updates reflect a form of maintenance where the model adapts to its task domain without major reconfiguration. As shown in Figure~\ref{fig:faid_drift_plot}, the model's trajectory validates the applicability of our drift detection pipeline and underscores the value of commit-level metric analysis in assessing semantic preservation.

\vspace{-0.8em}
\subsection{Summary of Findings}
Based on our empirical study, we identify several patterns of refactoring maintenance in Hugging Face model repositories: \vspace{-1em}
\begin{itemize}
    \item Automated metric extraction enables the identification of both positive drift (optimization) and semantic preservation patterns.\vspace{-0.5em}
    \item Commit histories coupled with public documentation provide underexplored yet valuable artifacts for refactoring analysis in ML.\vspace{-0.5em}
    \item Documentation quality significantly impacts observability of model evolution, indicating the need for community-wide standards. \vspace{-0.5em}
\end{itemize}
By applying automated metric extraction, we are able to empirically detect cases where key performance metrics remain stable across model updates, providing concrete evidence of semantic preservation in LESS.

\section{Conclusion}
\label{sec:conclusion}
As ML systems become increasingly modular and versioned, maintaining semantic preservation across model updates is critical for trust and traceability. In this paper, we present a scalable and interpretable pipeline for detecting semantic status in ML models using intra-repository metadata and documentation from the Hugging Face hub. This work contributes an empirical baseline to analyze semantic safety during ML model evolution. By combining structured filtering, metric extraction, and commit history analysis, our approach enables empirical evaluation of semantic preservation over time without requiring access to source code or internal training data.


We addressed \textbf{RQ1} as follows: We studied 536 models filtered through our framework from an original set of over a million models. Our work revealed that semantic drift can be effectively quantified through public documentation of these models. We also observed that performance optimization without radical model changes was possible in two out of three different domains. Addressing \textbf{RQ2}, we identified semantically preserved models whose accuracy remained stable across commits, as well as models with insufficient documentation to support traceable evaluation. These findings validate the hypothesis that intra-repository evolution offers rich signals for assessing behavioral changes in learning-enabled software systems (LESS).

\section{Limitations and Threats to Validity}
\label{sec:limit}
Our methodology relies on documentation quality and commit accessibility on Hugging Face. While we filtered out empty or gated models, authors vary in their use of structured evaluation metadata. Inconsistent use of \texttt{README.md} files and varied metric formats introduce noise, despite our use of regex extraction.

While high-level aggregate metrics provide a useful starting point, they can disguise bias and insufficiency at a whole system level. To enhance our understanding of semantic preservation, future work will expand beyond aggregate metrics. We will conduct more granular analyses, examining model predictions on specific data slices to detect subtle but important shifts in subgroup performance. This will provide a more complete picture of a model's behavior, ensuring its intended function is maintained for all user groups. We will also include other dimensions such as decision boundary shifts and subgroup performance changes.
Finally, we also plan to extend this framework to explore fairness~\parencite{nguyen_fix_2023}, efficiency, and other multi-objective dimensions~\parencite{jia_reless_2024} refactoring, while improving robustness across diverse documentation styles.
\label{sec:ack}

\section{Acknowledgments}

This material is based upon work supported by the National Science Foundation under Award Nos.~CCF-22-00343 and CCF-23-43750.



\printbibliography
\newpage
\appendix

\end{document}